\documentclass[%
superscriptaddress,
showpacs,preprintnumbers,
amsmath,amssymb,
aps,
prl,
longbibliography,
twocolumn
]{revtex4-1}
\usepackage{ulem}
\usepackage[dvipdfmx]{graphicx} 
\usepackage{dcolumn}
\usepackage{bm}
\usepackage{color}
\usepackage{xcolor}
\usepackage{multibib}
\usepackage[update,prepend]{epstopdf}

\usepackage{float}

\begin {document}

\title{Dimensional Crossover of Thermal Transport in Nanoconfined Liquids Driven by the Interplay of Quasi-One-Dimensional Structure and Wall Dissipation}

\author{Kenta Hisamoto}
\affiliation{Faculty of Mechanical Engineering, Kyoto Institute of Technology, Matsugasaki, Sakyo-ku, Kyoto 606-8585, Japan}

\author{Yusei Kobayashi}
\email{kobayashi@kit.ac.jp}
\affiliation{Faculty of Mechanical Engineering, Kyoto Institute of Technology, Matsugasaki, Sakyo-ku, Kyoto 606-8585, Japan}
\affiliation{High-Performance Simulation Research Center, Kyoto Institute of Technology, Matsugasaki, Sakyo-ku, 606-8585, Kyoto, Japan}

\author{Takahiro Ikeda}
\affiliation{High-Performance Simulation Research Center, Kyoto Institute of Technology, Matsugasaki, Sakyo-ku, 606-8585, Kyoto, Japan}
\affiliation{Center for the Possible Futures, Kyoto Institute of Technology, Matsugasaki, Sakyo-ku 606-8585, Kyoto, Japan}

\author{Eiji Yamamoto}
\affiliation{Department of System Design Engineering, Keio University, 3-14-1 Hiyoshi, Kohoku-ku, Yokohama, Kanagawa 223-8522, Japan}

\author{Masashi Yamakawa}
\affiliation{Faculty of Mechanical Engineering, Kyoto Institute of Technology, Matsugasaki, Sakyo-ku, Kyoto 606-8585, Japan}
\affiliation{High-Performance Simulation Research Center, Kyoto Institute of Technology, Matsugasaki, Sakyo-ku, 606-8585, Kyoto, Japan}

\begin{abstract}
Heat transport in nanoconfined liquids can deviate from ordinary Fourier behavior because confinement alters liquid structure and interfacial dissipation.
Although such changes may lead to quasi-one-dimensional transport or overdamped sound relaxation, the conditions under which length-dependent transport persists remain unclear.
Here we use molecular dynamics simulations of monatomic liquid argon confined in carbon nanotubes with systematically varied radii and lengths.
We find a radius-controlled crossover: length-dependent axial thermal conductivity persists over long tube lengths in single-file and single-shell states, but is strongly truncated or nearly saturated once mixed-shell or multilayer packing develops.
This crossover is accompanied by the loss of clear acoustic-like axial modes and enhanced wall--liquid friction.
Thus, tube radius controls whether length-dependent heat transport persists or is truncated by coupling confined-liquid structure to wall-induced dissipation.
\end{abstract}

\maketitle

Liquids under nanoconfinement can exhibit unique structural and transport properties that differ from those in the bulk.~\cite{Aluru:cr:2023,Li:jcp:2025}
In carbon nanotubes (CNTs) and related nanopores, confined water can form distinct low-dimensional structures, including ordered ice nanotubes, single-file chains, and density-dependent liquid phases.~\cite{Winarto:jcp:2015,Winarto:wat:2017,Koga:na:2001,Liu:an:2023,Mochizuki:pnas:2015,Nomura:pnas:2017}
Such low-dimensional structures can give rise to unusual thermal transport. For example, the thermal conductivity of water confined in quasi-one-dimensional nanotubes becomes length dependent in the single-file limit.~\cite{Zhao:jcp:2020,Imamura:jcp:2024}
Recent theoretical studies have further suggested that strong confinement can drive a crossover from ordinary three-dimensional transport to anomalous quasi-one-dimensional heat conduction.~\cite{luo:pre:2021,Luo:ent:2025}
Confinement can also suppress propagating sound modes, leading to overdamped or diffusive density relaxation.~\cite{Holey:prl:2023,Holey:prf:2024}
Such length-dependent heat transport is closely related to anomalous heat conduction in low-dimensional momentum-conserving solids and lattice models, where length-dependent thermal conductivity has been widely studied.~\cite{Lepri:pr:2003,Casati:pre:2003,Basile:prl:2006,Chang:prl:2008}

The liquid--solid interface can also influence transport in confined liquids.
Previous studies have shown that hydrodynamic boundary conditions, liquid--solid friction, and interfacial slip depend sensitively on surface chemistry and wettability, and that specific surface--liquid interactions can further modify interfacial friction.~\cite{Bocquet:pre:1994,Bocquet:csr:2010,Huang:pre:2014,Joly:jpcl:2016,Arai:nh:2023}
Thus, length-dependent heat transport in nanoconfined liquids should be affected not only by the structure and effective dimensionality of the confined liquid but also by dissipative coupling to the confining wall.
However, it remains unclear how the interplay between liquid structure and wall-induced dissipation controls the persistence or truncation of length-dependent heat transport.

Here, using molecular dynamics (MD) simulations, we show that the tube radius $R$ controls the crossover between persistent and truncated length-dependent heat transport in monatomic liquid argon confined in CNTs.
As $R$ increases from $0.4$ to $1.0\,{\rm nm}$, the confined liquid reorganizes from single-file and single-shell states to mixed-shell and multilayer packing, while the apparent axial thermal conductivity changes from strongly length dependent to nearly saturated.
By combining thermal-transport calculations with density-mode and wall-friction analyses, we identify this crossover as a consequence of coupled changes in confined-liquid structure and wall-induced dissipation.

\begin{figure*}[t]
	\centering
	\includegraphics[width=12.5cm]{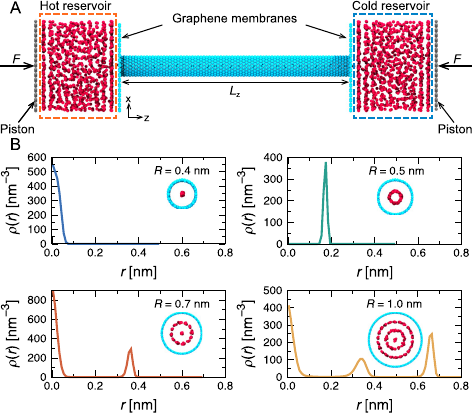}
	\caption{Simulation system and structure of confined argon. (A)~Representative snapshot of the CNT--reservoir system under a thermal gradient. Argon and carbon atoms are shown in red and light blue, respectively. (B)~Radial number density profiles of confined argon in CNTs with different radii, $R$, at $L_z=70\,{\rm nm}$. Representative cross-sectional snapshots are also shown.}
	\label{fig:system}
\end{figure*}

Figure~\ref{fig:system}A shows the simulation setup used to examine heat transport in liquid argon confined in CNTs.
A central CNT is connected to two argon reservoirs, and a temperature gradient is imposed along the tube axis.
We vary the tube radius from $R=0.4$ to $1.0\,{\rm nm}$ and the tube length from $L_z=1$ to $300\,{\rm nm}$.
Argon--argon and argon--carbon interactions are described by Lennard--Jones potentials, and the tube wall atoms are harmonically restrained.
Further simulation details are provided in Sec.~S1 of the Supporting Information.

Before analyzing thermal transport, we first characterize how the confined-liquid structure depends on the tube radius.
Figure~\ref{fig:system}B shows radial number density profiles, $\rho(r)$, together with representative cross-sectional snapshots for $R=0.4$, $0.5$, $0.7$, and $1.0\,{\rm nm}$ at $L_z=70\,{\rm nm}$.
The same structural features are observed over the range of tube lengths examined, indicating that the radius, rather than the length, primarily determines the radial packing state.
At $R=0.4\,{\rm nm}$, $\rho(r)$ exhibits a single central peak, corresponding to a quasi-one-dimensional single-file structure along the tube axis.
At $R=0.5\,{\rm nm}$, the density maximum shifts away from the center to $r\approx 0.17\,{\rm nm}$, indicating the formation of a single-shell structure.
At $R=0.7\,{\rm nm}$, $\rho(r)$ shows both a central peak and an outer peak at $r\approx 0.36\,{\rm nm}$, consistent with a mixed-shell structure composed of a single-file-like core and a surrounding shell.
At $R=1.0\,{\rm nm}$, three peaks appear at approximately $r\approx 0$, $0.36$, and $0.66\,{\rm nm}$, indicating multilayer packing.
The cross-sectional snapshots provide direct visual confirmation of these radius-dependent packing motifs.
The average number density remains nearly constant after equilibration for all radii (Fig.~S1), indicating that the observed structures are stable over the analyzed time window.
Thus, changing $R$ reorganizes the confined liquid from a quasi-one-dimensional single-file state to single-shell, mixed-shell, and multilayer packing states, rather than simply increasing the available cross-sectional area.

\begin{figure}[tb]
	\centering
	\includegraphics[width=8.5cm]{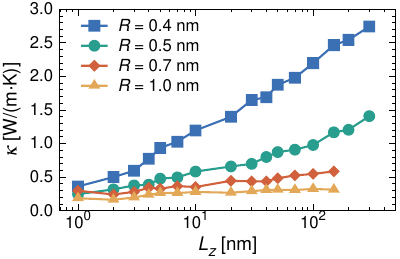}
	\caption{Thermal conductivity, $\kappa$, as a function of nanotube length, $L_z$, for different tube radii, $R$.}
	\label{fig:kappa}
\end{figure}
\begin{figure*}[t]
	\centering
	\includegraphics[width=15.5cm]{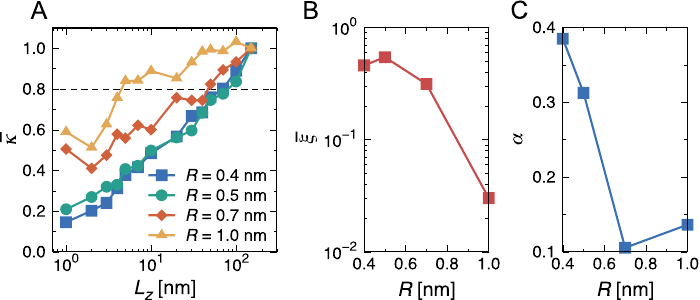}
	\caption{Quantification of the thermal transport crossover. (A)~Normalized thermal conductivity, $\bar{\kappa}=\kappa(L_z,R)/\kappa(L_z^\ast,R)$ with $L_z^\ast=150\,{\rm nm}$, as a function of nanotube length $L_z$ for different radii $R$. (B)~Normalized operational crossover length, $\bar{\xi}=\xi/L_z^\ast$, defined by $\bar{\kappa}(\xi,R)=0.8$. (C)~Effective growth exponent, $\alpha$, obtained from a log--log fit of $\kappa$ over $L_z=10$--$100\,{\rm nm}$ (Fig.~S2).}
	\label{fig:crossover}
\end{figure*}
We next examine whether this radius-dependent structural reorganization is reflected in thermal transport.
Figure~\ref{fig:kappa} shows the axial thermal conductivity, $\kappa$, as a function of nanotube length, $L_z$, for each $R$.
As a reference, the bulk thermal conductivity of liquid argon at $T=112.5\,{\rm K}$ and $P=3.0\,{\rm MPa}$ is $\kappa_{\rm bulk}=0.08\pm0.001\,{\rm W/(m\cdot K)}$.
The length dependence of $\kappa$ changes systematically with tube radius.
For $R=0.4\,{\rm nm}$, where the liquid forms a quasi-one-dimensional single-file structure, $\kappa$ increases strongly over the entire range examined, from $L_z=1.0$ to $300\,{\rm nm}$.
For $R=0.5\,{\rm nm}$, corresponding to a single-shell structure, a similar but weaker increase is observed.
By contrast, the growth of $\kappa$ is strongly suppressed for $R=0.7\,{\rm nm}$, where a mixed-shell structure forms, and is nearly absent for $R=1.0\,{\rm nm}$, where multilayer packing appears.
Thus, the persistence of length-dependent thermal transport is most pronounced in the single-file and single-shell states and is strongly reduced once mixed-shell or multilayer packing develops.

To quantify how far the length-dependent increase of $\kappa$ persists for different radii, we introduce an operational crossover length, $\xi$.
For each radius, we first normalize the conductivity as
$\bar{\kappa}=\kappa(L_z,R)/\kappa(L_z^{\ast},R)$, where $L_z^{\ast}=150\,{\rm nm}$ is the longest tube length common to all radii.
We define $\xi$ as the tube length at which $\bar{\kappa}(L_z,R)$ reaches 0.8 upon increasing $L_z$, and use the normalized crossover length $\bar{\xi}=\xi/L_z^{\ast}$.
A larger $\bar{\xi}$ indicates that the length-dependent growth of $\kappa$ persists to longer tube lengths, whereas a smaller $\bar{\xi}$ indicates earlier truncation of the length dependence.
Thus, $\bar{\xi}$ provides an operational measure of the axial length scale over which length-dependent heat transport persists, rather than a sharp physical transition length.
Figure~\ref{fig:crossover}A,B shows $\bar{\kappa}$ and the resulting radius dependence of $\bar{\xi}$.
The normalized crossover length decreases from $\bar{\xi}\approx 4.5\times10^{-1}$ for $R=0.4$ and $0.5\,{\rm nm}$ to $\bar{\xi}\approx 3.0\times10^{-1}$ for $R=0.7\,{\rm nm}$ and $\bar{\xi}\approx 3.0\times10^{-2}$ for $R=1.0\,{\rm nm}$.
Thus, the operational length scale over which length-dependent heat transport persists is shortened by more than one order of magnitude as the packing changes from single-file or single-shell states to mixed-shell or multilayer states.

We further quantify the length dependence of $\kappa$ by evaluating an effective growth exponent, $\alpha$.
The exponent was obtained from the slope of a log--log plot of $\kappa$ versus $L_z$ over $L_z=10$--$100\,{\rm nm}$ (Fig.~S2), corresponding to an apparent scaling relation $\kappa\sim L_z^{\alpha}$ in this length range.
As shown in Fig.~\ref{fig:crossover}C, $\alpha$ decreases from $\alpha\approx 0.38$ at $R=0.4\,{\rm nm}$ and $\alpha\approx 0.31$ at $R=0.5\,{\rm nm}$ to $\alpha\approx 0.10$ at $R=0.7\,{\rm nm}$, with only a slight increase to $\alpha\approx 0.13$ at $R=1.0\,{\rm nm}$.
Thus, the effective exponent is substantially larger in the single-file and single-shell states than in the mixed-shell and multilayer states, indicating that the radius-induced structural crossover reduces the apparent power-law growth of $\kappa(L_z)$.

In low-dimensional momentum-conserving systems, length-dependent heat conduction is commonly attributed to long-lived hydrodynamic modes, such as sound modes and the heat mode.~\cite{Lepri:pr:2003,Dhar:ap:2008,Narayan:prl:2002}
In confined fluids, however, wall-induced dissipation can strongly damp sound modes and drive their relaxation toward overdamped or diffusion-dominated behavior.~\cite{Holey:prl:2023,Holey:prf:2024}
To examine whether the persistence of $\kappa(L_z)$ is related to the propagation and relaxation of axial density modes, we performed a mode-resolved analysis of axial density fluctuations (Fig.~S3).
For $R=0.4$ and $0.5\,{\rm nm}$, the peak frequencies increase approximately linearly with $k_n$ at low wave numbers, allowing $c_{\rm s}$ to be estimated from the slope of the dispersion relation (Fig.~S4).
We then estimated an acoustic propagation length, $c_{\rm s}\tau^{\ast}$, where $\tau^{\ast}$ was obtained from the linewidth of the lowest-order axial density mode (Table~S1).
For $R=0.7$ and $1.0\,{\rm nm}$, the spectra still show finite-frequency peaks, but the peaks are broadened and do not yield a clear linear low-$k$ dispersion.
Thus, a robust sound speed, acoustic lifetime, and acoustic propagation length were not assigned for these larger-radius tubes.
The resulting values of $c_{\rm s}\tau^{\ast}$ for $R=0.4$ and $0.5\,{\rm nm}$ are nearly identical.
Thus, the radius-dependent crossover in $\kappa(L_z)$ cannot be organized by a simple acoustic propagation length alone.
\begin{figure}[t]
	\centering
	\includegraphics[width=8.5cm]{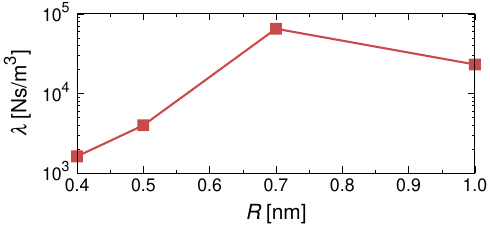}
	\caption{Wall--liquid friction coefficient, $\lambda$, as a function of tube radius $R$ at $L_z=70\,{\rm nm}$. Values were obtained from the Green--Kubo integral of the axial force autocorrelation.}
	\label{fig:lambda}
\end{figure}

We next examine boundary dissipation as an indicator of transport truncation, quantified by the wall--liquid friction coefficient, $\lambda$.
In equilibrium molecular dynamics simulations, $\lambda$ was evaluated using the Green--Kubo relation,~\cite{Bocquet:pre:1994,Bocquet:fd:1999}
\begin{equation}
    \lambda = \frac{1}{A k_{\rm B} T}\int_0^\infty \langle F_{z}(0) F_{z}(t) \rangle dt,
\end{equation}
where $A$ is the CNT surface area and $F_{z}$ is the axial force exerted by the confined liquid on the wall.
The force autocorrelation functions are shown in Fig.~S5, and details are given in Sec.~S5 of the Supporting Information.
As shown in Fig.~\ref{fig:lambda}, $\lambda$ increases only modestly from $R=0.4$ to $0.5\,{\rm nm}$, then rises sharply at $R=0.7\,{\rm nm}$ by more than an order of magnitude.
Although $\lambda$ decreases somewhat at $R=1.0\,{\rm nm}$, it remains substantially larger than at $R=0.4$ and $0.5\,{\rm nm}$.
Systems with pronounced $L_{z}$ dependence in $\kappa$ therefore correspond to a low-friction regime, whereas those with weak or saturated growth fall into a high-friction regime.
This correspondence suggests that weaker wall dissipation is associated with more persistent length-dependent growth of $\kappa$, whereas stronger wall coupling is associated with earlier truncation of $\kappa(L_z)$.

Although larger $\lambda$ values are associated with suppressed or saturated growth of $\kappa(L_z)$, wall friction alone does not provide a complete description of the transport crossover.
As shown in Fig.~\ref{fig:system}B, varying $R$ also reorganizes the confined-liquid structure, from single-file and single-shell states at smaller radii to mixed-shell and multilayer packing at larger radii.
This structural reorganization changes not only the packing geometry, but also the character of axial density fluctuations and their dissipative coupling to the wall.
The radius-dependent transport regimes can therefore be understood as the combined result of confined-liquid structure, collective axial density modes, and boundary dissipation.
At $R=0.4\,{\rm nm}$, the liquid forms a quasi-one-dimensional single-file structure, remains in the low-friction regime, and exhibits the most pronounced growth of $\kappa(L_z)$.
At $R=0.5\,{\rm nm}$, the liquid adopts a single-shell structure and still exhibits appreciable length-dependent growth.
In contrast, at $R=0.7$ and $1.0\,{\rm nm}$, mixed-shell or multilayer packing is accompanied by broadened axial spectra and substantially larger wall friction, and the growth of $\kappa(L_z)$ is strongly truncated.
Thus, the tube radius controls thermal transport not solely as a geometric parameter, but through its coupled effects on confined-liquid structure, collective density fluctuations, and wall-induced dissipation.

In summary, we have shown that liquid argon confined in carbon nanotubes exhibits a radius-controlled crossover in length-dependent thermal transport.
For smaller tubes of $R=0.4$ and $0.5\,{\rm nm}$, $\kappa(L_z)$ grows strongly with tube length, whereas for larger tubes of $R=0.7$ and $1.0\,{\rm nm}$ this growth is strongly truncated or nearly saturated.
This crossover shows that radial confinement does not merely change the magnitude of $\kappa$, but determines whether length-dependent heat transport persists over long axial distances or is truncated at much shorter lengths.
Mode-resolved density-fluctuation analysis shows that the smaller-radius tubes retain clearer acoustic-like axial modes, whereas the larger-radius tubes exhibit broadened spectra from which a robust acoustic propagation length cannot be assigned.
Instead of being organized by acoustic propagation length alone, the crossover emerges from the coupling between confined-liquid structure and wall-induced dissipation.
These results identify tube radius as a control variable linking confined-liquid structure, boundary dissipation, and the persistence of length-dependent heat transport.

\begin{acknowledgments}
Y.K. acknowledges JSPS KAKENHI Grant No.~JP24K17216 and the support of KIT Grants-in-Aid for Early-Career Scientists. T.I. was supported by JSPS KAKENHI Grant No.~JP25K23425.
\end{acknowledgments}


\bibliography{cite} 


\clearpage

\onecolumngrid
{
    \center \bf \large 
    Supplementary Materials for ``Dimensional Crossover of Thermal Transport in Nanoconfined Liquids Driven by the Interplay of Quasi-One-Dimensional Structure and Wall Dissipation''\vspace*{1cm}\\ 
    \vspace*{0.0cm}
}
\twocolumngrid  



\setcounter{figure}{0}


\section*{Simulation details}
Simulation system consisted of a central carbon nanotube (CNT) region connected to two argon reservoirs located at its left and right ends (see Fig.~\ref{fig:system}A of the main text). Here, argon was chosen as a minimal monatomic liquid model to isolate the effect of nanoconfinement on thermal transport. Because argon is electrically neutral and does not exhibit directional interactions such as hydrogen bonding, both the intermolecular and wall--liquid interactions can be described in a simple and well-defined manner by the Lennard--Jones (LJ) potential. This simplicity allows the present study to focus on how the tube radius reorganizes the confined-liquid structure and modifies wall-induced dissipation, without additional complications arising from molecular orientation, charge distribution, or association effects. Argon is therefore used here not to represent a specific application fluid, but as a reference system for identifying the generic physical mechanisms governing heat transport under nanoconfinement. The CNT and the reservoirs were separated by graphene membranes, and movable pistons were placed outside the reservoirs to maintain the liquid state of argon under confinement. The CNT had a circular cross section with radius $R$ and length $L_{z}$, and the geometric anisotropy of the system was controlled by varying both $R$ and $L_{z}$. To ensure compatibility of the graphene sheets with the periodic boundaries, the lateral box lengths are fixed at $L_{x}=3.829\,{\rm nm}$ and $L_{y}=3.684\,{\rm nm}$ for all systems. The CNT wall atoms are tethered to their initial lattice positions through a harmonic spring potential, $U_{\rm w}=k_{\rm w}(r-r_{\rm 0})^{2}$, where $k_{\rm w}=69.5\,{\rm N/m}$ denotes the wall stiffness~\cite{Nordquist:jpcb:2022,Huda:jpcb:2022,Lobato:jms:2023} and $r-r_{\rm 0}$ is the displacement from the initial position.~\cite{Kim:mn:2008} Nonbonded interactions between argon atoms and between argon and carbon atoms are described by the LJ potential,
\begin{equation}
	U_{\rm LJ}(r_{ij}) = \begin{cases}
	4\varepsilon_{ij} \left[\left(\frac{\sigma_{ij}}{r_{ij}}\right)^{12}-\left(\frac{\sigma_{ij}}{r_{ij}}\right)^{6}\right],
	\;\;\;\; r_{ij} \leq r_{\rm c} \\
	0, \;\;\;\;\;\;\;\;\;\;\;\;\;\;\;\;\;\;\;\;\;\;\;\;\;\;\;\;\;\;\;\;\;\;\;\;\;\;\;\; r_{ij}>r_{\rm c},
	\end{cases}
	\label{eq:ULJ}
\end{equation}
where $r_{ij}$ is the distance between the $i$-th and $j$-th particles, $\varepsilon_{ij}$ is interaction strength, and $\sigma_{ij}$ is the distance parameter. A cutoff distance $r_{\rm c}$ of $1.5\,{\rm nm}$ is used for all LJ interactions. The LJ parameters for argon--argon and carbon--carbon interactions are set to $\varepsilon_{\rm ar}=9.96\times 10^{2}\,{\rm J/mol}$ and $\sigma_{\rm ar}=0.3405\,{\rm nm}$ and $\varepsilon_{\rm c}=3.61\times 10^{2}\,{\rm J/mol}$ and $\sigma_{\rm c}=0.3400\,{\rm nm}$, respectively, and the Lorentz--Berthelot mixing rules~\cite{allen:book:1989} are applied to describe the interactions between unlike pairs.

Argon atoms initially placed in the two reservoirs were driven into the CNT by the pistons during equilibration at $T=112.5\,{\rm K}$. During this stage, a Nos\'{e}--Hoover thermostat~\cite{Nose:ptrs:1991} was applied together with a piston pressure of $P_{zz}=3.0\,{\rm MPa}$ along the $z$ direction so that argon remained in the liquid state under confinement. After confirming that the CNT was fully filled with argon, we carried out additional equilibrium simulations for $5$--$10\,{\rm ns}$. A temperature gradient was then imposed by thermostating the left and right reservoirs at $130$ and $95\,{\rm K}$, respectively, using the canonical sampling through velocity rescaling method.~\cite{Bussi:jcp:2007} The nonequilibrium simulations were continued for at least $10\,{\rm ns}$ until a steady state was established, and data were subsequently collected for $10$--$20$ ns after the temperature profile and heat flux had become stationary. The equations of motion were integrated with a time step of $2\,{\rm fs}$ throughout all stages of the simulations. All simulations were performed using GPU-accelerated versions~\cite{Brown:cpc:2011} of the Large-scale Atomic/Molecular Massively Parallel Simulator (LAMMPS).~\cite{Plimpton:jcp:1995}

The thermal conductivity of liquids within the CNT in the $z$ direction was evaluated from Fourier's law,
\begin{equation}
	\kappa = -\frac{J_z}{dT/dz},
\end{equation}
where $J_{z}$ is the steady heat flux in the $z$ direction and $dT/dz$ is the corresponding temperature gradient inside the CNT. The heat flux $J_{z}$ was calculated from the Irving--Kirkwood expression~\cite{Irving:jcp:1950,Kjelstrup:book:2008},
\begin{eqnarray}
	J_z &=&\frac{1}{V_{\rm tube}} \nonumber \\
	& & \left[\sum_{i \in V} e_i v_{i,z}+\frac{1}{2}\sum_{i<j \in V}
	\left\{\mathbf{F}_{ij} \cdot \left(\mathbf{v}_i+\mathbf{v}_j \right)\right\}r_{ij,z}\right],
	\label{eq:heat_flux}
\end{eqnarray}
where $e_{i}$ and $v_{i,z}$ are the energy and axial velocity of atom $i$, respectively, $F_{ij}$ is the interatomic force between atoms $i$ and $j$, $r_{ij,z}$ is their axial distance, and $V_{\rm tube}$ is the analysis volume of the CNT region. Here, $V_{\rm tube}$ was taken as $V=\pi R^{2}L^{'}$, where $L^{'}$ is the effective analysis length in the $z$ direction. To avoid artifacts associated with the CNT--reservoir junctions, atoms in the vicinity of the junction regions are excluded from the analysis, so that $L^{'}$ was shorter than the actual tube length $L_{z}$. The temperature profile was then evaluated along the $z$-axis, and $\nabla T$ was obtained from the approximately linear region inside the CNT after excluding the junction-adjacent parts where the local temperature exhibited stronger fluctuations, based on a previous study.~\cite{Imamura:jcp:2024}

\section*{Time evolution of the number density}
\begin{figure*}[t]
	\centering
	\includegraphics[width=7.5cm]{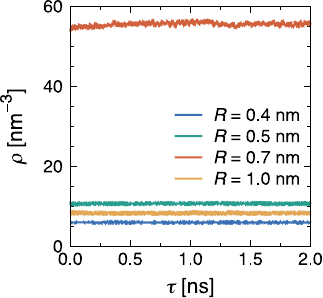}
	\caption{Time evolution of the number density, $\rho$, of the confined argon for various nanotube radii, $R$ at $L_{z}=70\,{\rm nm}$.}
	\label{fig:num_dens}
\end{figure*}

To confirm that the amount of argon confined inside the CNT remained stationary during the simulation, we monitored the time evolution of the number density. At each sampled time, the instantaneous number density was calculated as
\begin{equation}
	\rho(t)=\frac{N_{\rm ar}(t)}{V_{\rm tube}},
	\label{eq:number_density}
\end{equation}
where $N_{\rm ar}(t)$ is the number of argon atoms inside the CNT region and $V_{\rm tube}$ is the tube volume. Figure~\ref{fig:num_dens} shows that the number density exhibits only small fluctuations around a constant mean value for all $R$, indicating that the filling state of the confined liquid is stable over the analyzed time window.

\section*{Estimation of the effective growth exponent}
\begin{figure*}[b]
	\centering
	\includegraphics[width=7.5cm]{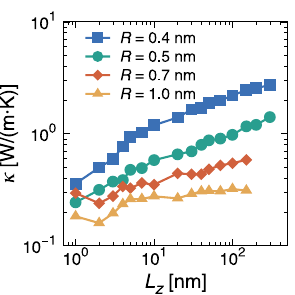}
	\caption{Thermal conductivity, $\kappa$, as a function of nanotube length, $L_{z}$, for various tube radii, $R$, on a log--log scale. The effective growth exponent $\alpha$ shown in Fig.~3C of the main text was obtained from the slope over $L_{z}=10$--$100\,{\rm nm}$.}
	\label{fig:kappa_loglog}
\end{figure*}

The effective growth exponent, $\alpha$, was evaluated from the length dependence of the apparent thermal conductivity. Figure~\ref{fig:kappa_loglog} shows the corresponding log--log plots of $\kappa$ versus $L_{z}$. The extracted $\alpha$ values are shown in Fig.~3C of the main text and are used as a complementary measure of the strength of conductivity growth for different tube radii.

\section*{Density-mode analysis and acoustic reference length}
\begin{figure*}[t]
	\centering
	\includegraphics[width=14.5cm]{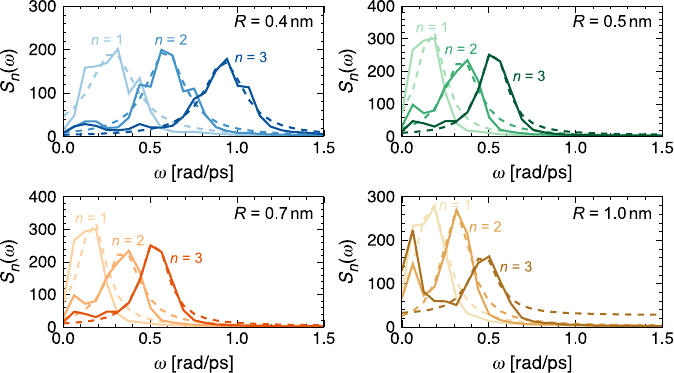}
	\caption{Representative spectra $S_{n}(\omega)$ of the axial density modes at $L_{z}=70\,{\rm nm}$ for various $R$ as indicated. Solid lines show the spectra for the lowest-order modes ($n=1$--$3$), and dashed lines indicate the corresponding Lorentzian fits used to estimate the Brillouin peak frequency $\omega_{n}$ and linewidth $\Gamma_{n}$.}
	\label{fig:Sn_omega}
\end{figure*}

\begin{figure*}[b]
	\centering
	\includegraphics[width=14.5cm]{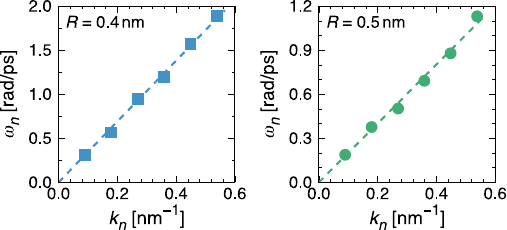}
	\caption{Dispersion relation of the axial density modes for $R=0.4$ and $0.5\,{\rm nm}$ at $L_{z}=70\,{\rm nm}$. Symbols show the Brillouin peak frequencies $\omega_{n}$ extracted from the spectra of the lowest-order axial density modes, plotted against the corresponding wave numbers $k_{n}=2\pi n/L_{z}$. Dashed lines indicate linear fits, $\omega_{n}=c_{\rm s}k_{n}$, from which the sound speed $c_{\rm s}$ was estimated.}
	\label{fig:omega_k}
\end{figure*}

To assess whether the crossover in $\kappa(L_{z})$ can be explained by a simple acoustic propagation length, we analyzed equilibrium density fluctuations along the $z$-axis at $L_{z}=70\,{\rm nm}$. For a given tube length $L_{z}$, the axial density mode was defined as
\begin{equation}
	\rho_{k_n}(t)=\sum_{j\in {\rm Ar}}\exp\left[-i k_n z_j(t)\right],\qquad
	k_n=\frac{2\pi n}{L_z},
\label{eq:rho_kn}
\end{equation}
where $z_{j}(t)$ is the axial coordinate of atom $j$ and $n=1$, $2$, $\ldots$ labels the longitudinal mode. From the time series of $\rho_{k_n}(t)$, we calculated the normalized autocorrelation function
\begin{eqnarray}
	&& C_n(t)=\frac{\left\langle\delta\rho_{k_n}^{*}(t_0)\,\delta\rho_{k_n}(t_0+t)\right\rangle_{t_0}}{\left\langle
\left|\delta\rho_{k_n}(t_0)\right|^2\right\rangle_{t_0}},\qquad \nonumber \\
	&& \delta\rho_{k_n}(t)=\rho_{k_n}(t)-\left\langle \rho_{k_n} \right\rangle .
	\label{eq:rho_autocorr}
\end{eqnarray}
Here, the asterisk denotes complex conjugation, and $\langle \cdots \rangle$ denotes an average over multiple time origins. The corresponding spectrum is then obtained from the Fourier transform of $C_{n}(t)$, and the Brillouin peak frequency $\omega_{n}$ was identified from the positive-frequency peak of the real spectral component after excluding the zero-frequency contribution. Representative spectra are shown in Fig.~\ref{fig:Sn_omega}. The sound speed, $c_{\rm s}$, was estimated from the low-$k$ dispersion relation
\begin{equation}
	\omega_n \simeq c_{\rm s} k_n .
	\label{eq:sound_dispersion}
\end{equation}
using the lowest accessible longitudinal modes. The resulting $\omega_{n}$ {\it vs.} $k_{n}$ plots are shown in Fig.~\ref{fig:omega_k}. To estimate the lifetime of the propagating density fluctuation, the Brillouin peak is fitted by a Lorentzian form~\cite{Schoen:mp:1986,Porcheron:pre:2002,Brandt:bj:2009},
\begin{equation}
	S_n(\omega)\simeq\frac{A_n \Gamma_n^2}{(\omega-\omega_n)^2+\Gamma_n^2}+B_n ,
	\label{eq:lorentzian_fit}
\end{equation}
where $\Gamma_{n}$ is the half-width at half maximum. A characteristic relaxation time is then defined as $\tau_{n}=\Gamma_{n}^{-1}$. In this study, we use the representative relaxation time of the lowest-order ({\it i.e.}, longest-wavelength) axial density mode, $\tau^{\ast} \equiv \tau_{1}$. This mode is the most relevant to the longest-range propagating correlation accessible in a given tube. The corresponding acoustic reference length was then defined as
\begin{equation}
	\ell(R)=c_{\rm s}(R)\tau^{\ast}(R).
	\label{eq:acoustic_length}
\end{equation}
Table~\ref{tab:Extracted} summarizes the extracted values of $c_{\rm s}$, $\tau^{\ast}$, and $\ell(R)$. For $R=0.7$ and $1.0\,{\rm nm}$, a clear monotonic dispersion of the spectral peak with $k_{n}$ was not obtained. Therefore, a robust sound speed and acoustic lifetime were not extracted from the mode-resolved analysis.

\section*{Wall--liquid friction coefficient}
\begin{figure*}[b]
	\centering
	\includegraphics[width=13.5cm]{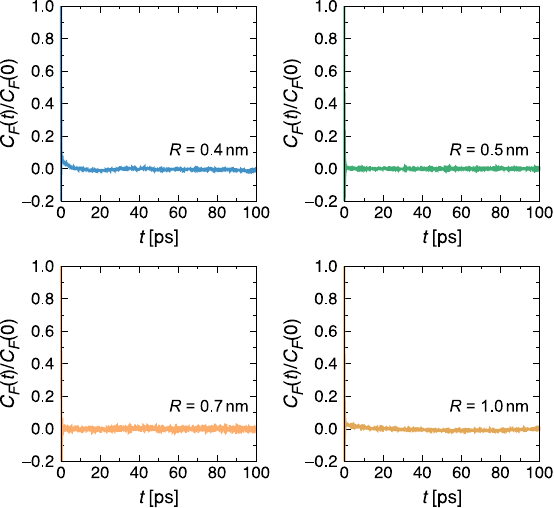}
	\caption{Normalized force autocorrelation functions, $C_{F}(t)/C_{F}(0)$, for each CNT radius, $R$ at $L_{z}=70\,{\rm nm}$. Here, $C_{F}(t)=\left\langle F_{z}(0)F_{z}(t)\right\rangle$, where $F_{z}$ is the instantaneous total axial force exerted by the confined liquid on the CNT wall.}
	\label{fig:friction}
\end{figure*}

To quantify boundary dissipation, we evaluated the wall--liquid friction coefficient, $\lambda$, from equilibrium fluctuations of the tangential force exerted by the confined liquid on the CNT wall. In equilibrium MD simulations, $\lambda$ was evaluated using the Green--Kubo relation
\begin{equation}
	\lambda=\frac{1}{A k_{\rm B} T}\int_{0}^{\infty}\left\langle F_z(0)F_z(t)\right\rangle\,dt ,
	\label{eq:friction_coefficient}
\end{equation}
where $A$ is the inner surface area of the CNT, $k_{\rm B}$ is the Boltzmann constant, $T$ is the temperature, and $F_{z}(t)$ is the instantaneous total force in the $z$-axis exerted by the confined liquid on the CNT wall. Figure~\ref{fig:friction} shows the normalized force autocorrelation functions, $C_{F}(t)/C_{F}(0)$ at $L_{z}=70\,{\rm nm}$. Here, $C_{F}(t)=\left\langle F_{z}(0)F_{z}(t)\right\rangle$, where $F_{z}$ is the instantaneous total axial force exerted by the confined liquid on the CNT wall. Because Green--Kubo integrals for liquid--solid friction in finite-size simulations do not necessarily exhibit a true long-time plateau, a practical finite-time cutoff is commonly required. Previous studies have therefore used, for example, the maximum of the Green--Kubo integral or the value of the integral at the first zero of the corresponding force autocorrelation function as practical estimators of the friction coefficient.~\cite{Brey:jcp:1982,Bocquet:pre:1994,Barrat:mp:2003,Joly:jpcl:2016,Nakaoka:jcp:2017,Oga:jcp:2019} In the present work, we adopted a first-zero criterion based on the ensemble-averaged force autocorrelation function. To implement this criterion robustly for discretely sampled noisy data, we additionally required that, after the first positive-to-nonpositive crossing beyond a minimum lag time, the subsequent several samples remained nonpositive.

\clearpage
\begin{table*}[t]
\begin{center}
\caption{Extracted acoustic quantities.}
	
	\small
		\begin{tabular*}{0.9\textwidth}{@{\extracolsep{\fill}}cccc}
		\hline
		 $R$\,[nm]   & $c_{\rm s}$\,[m/s] & $\tau^{*}$\,[ps]   & $\ell(R)=c_{\rm s}\tau^{*}$\,[nm]  \\
		\hline
		 0.4         & 3454          & 6.3              & 21.8  \\
		 0.5         & 2008          & 10.9             & 21.9  \\
		 0.7         & ---           & ---              & ---  \\
		 1.0        & ---           & ---              & ---  \\
		\hline
		\end{tabular*}
	\label{tab:Extracted}
\end{center}
\end{table*}


\end{document}